\documentclass[12pt]{article}
\usepackage{a4wide}
\usepackage{amssymb}
\usepackage{amsmath}
\usepackage{graphicx}
\usepackage{mathdots}
\usepackage{slashed}
\usepackage{verbatim}
\usepackage{xcolor}

\usepackage{hyperref}

\usepackage{IEEEtrantools}

\def\makeatletter{\catcode`\@=11}
\makeatletter
\def\mathbox#1{\hbox{$\m@th#1$}}%
\def\math@ccstyles#1#2#3#4#5#6#7{{\leavevmode
      \setbox0\mathbox{#6#7}%
      \setbox2\mathbox{#4#5}%
      \dimen@ #3%
      \baselineskip\z@\lineskiplimit#1\lineskip\z@
      \vbox{\ialign{##\crcr
             \hfil \kern #2\box2 \hfil\crcr
             \noalign{\kern\dimen@}%
             \hfil\box0\hfil\crcr}}}}
\def\mathaccstyles{\math@ccstyles\maxdimen}
\def\maththroughstyles{\math@ccstyles{-\maxdimen}}
\def\unity%
 {\maththroughstyles{.45\ht0}\z@\displaystyle {\mathchar"006C}\displaystyle 1}
 
\begin{document}

\mbox{}
\vspace{0truecm}
\linespread{1.1}


\centerline{\LARGE \bf A Scaling Limit for Line and Surface Defects}
%

\medskip

\vspace{.4cm}

 \centerline{\LARGE \bf }

\vspace{1.5truecm}

\centerline{
    { \bf D. Rodriguez-Gomez${}^{a,b}$} \footnote{d.rodriguez.gomez@uniovi.es}}

\vspace{1cm}
\medskip
\centerline{{\it ${}^a$ Department of Physics, Universidad de Oviedo}} \centerline{{\it C/ Federico Garc\'ia Lorca  18, 33007  Oviedo, Spain}}
\medskip
\centerline{{\it ${}^b$  Instituto Universitario de Ciencias y Tecnolog\'ias Espaciales de Asturias (ICTEA)}}\centerline{{\it C/~de la Independencia 13, 33004 Oviedo, Spain.}}
\vspace{1cm}

\centerline{\bf ABSTRACT}
\medskip 

We study symmetry-breaking line defects in the Wilson-Fisher theory with $O(2N+1)$ global symmetry near four dimensions and symmetry-preserving surface defects in a cubic model with $O(2N)$ global symmetry near six dimensions. We introduce a scaling limit inspired by the large charge expansion in Conformal Field Theory. Using this, we compute the beta function for the defect coupling which allows to identify the corresponding Defect Conformal Field Theories. We also compute the correlation function of two parallel defects as well as correlation functions of certain defect operators with large charge under the surviving symmetry.

\noindent

\newpage
\setcounter{footnote}{0}

\tableofcontents

\section{Introduction and conclusions}

The study of defects in Quantum Field Theory --and more specifically in Conformal Field Theory (CFT)--  is very interesting for a number of reasons. For instance, in realistic Condensed Matter scenarios, defects are often present in the form of impurities. From a more theoretical point of view, defects in a given bulk CFT allow to probe interesting Physics which has recently attracted much attention, including the discovery of new central charges and the study properties of RG flows (see \textit{e.g} \cite{Herzog:2019rke,Bianchi:2019umv,Herzog:2021hri,Bianchi:2021snj,Cuomo:2021rkm,Chalabi:2021jud,Cuomo:2021kfm,Cuomo:2022xgw} for a very partial list of some of the most recent developments). Moreover, in gauge theories, defects (or equivalently, extended operators) play a very important role in understanding central aspects including the symmetries and phases of a given theory (see \textit{e.g.} \cite{Aharony:2013hda,Gaiotto:2014kfa}).

When coupled to a CFT, a defect triggers a Renormalization Group (RG) flow which in many cases ends in a fixed point. This fixed point of the theory hosted on the defect defines a Defect Conformal Field Theory (DCFT). In this work we will be interested in the simplest incarnation of this idea in the context of the Wilson-Fisher theory near $d=4$ and $d=6$. We consider a trivial defect in the UV and turn on a certain perturbation which triggers an RG flow ending in a DCFT. The particular class of perturbation that we will consider is a relative of the ``pinning field defect" studied recently in \cite{Cuomo:2021kfm,Cuomo:2022xgw} (see also \cite{Allais:2014fqa}).

The bulk theory which we consider is the Wilson-Fisher (WF) theory with $O(M)$ global symmetry near $d=4, 6$ ($M=2N+1$ near $d=4$ and $M=2N$ near $d=6$). In $d=4-\epsilon$ dimensions, the WF theory flows to an IR fixed point where the coupling is proportional to $\epsilon$. Formally, this fixed point can be continuated into a UV fixed point at negative $\epsilon$ --that is, to $d=4+|\epsilon|$. It has been argued in \cite{Fei:2014yja} that the appropriate UV completion is a certain cubic theory whose IR fixed point in $d=6-\epsilon$ dimensions concides with the UV fixed point of the quartic theory. Nevertheless the fact that the cubic potential is not bounded from below (or, alternatively, that the quartic theory at the UV fixed point has negative coupling) ends up manifesting itself through small imaginary parts in the anomalous dimensions, showing that the theory is best thought of as complex CFT \cite{Giombi:2019upv,Giombi:2020enj}. To these bulk theories we consider coupling a trivial defect, supported in a line in the case of $d=4-\epsilon$ and in a 2d surface in the case of $d=6-\epsilon$. In both cases, we deform by an operator inserted on the defect which is linear in a bulk field. This triggers in both cases an RG flow ending on a fixed point. There is a big difference though in that in the $d=4-\epsilon$ case the defect breaks global $O(2N+1)$ symmetry down to $O(2N)$ while in the $d=6-\epsilon$ case it does not break the $O(2N)$ symmetry.

Along an \textit{a priori} unrelated line, over the recent past a new manifestation of the well-known observation that ``large quantum numbers simplify things" has bee found. In particular, it has been realized that the sector of operators of large charge under a global symmetry in a CFT can be accessed regardless on the coupling $g$ through a new semiclassical approximation \cite{Hellerman:2015nra,Alvarez-Gaume:2016vff,Hellerman:2017sur} (see \cite{Gaume:2020bmp} for a review).  In the context of the Wilson-Fisher model studied in this paper, the emergence of such new semiclassical expansion can be understood ``microscopically" from the path integral \cite{Arias-Tamargo:2019xld,Badel:2019oxl,Watanabe:2019pdh} (see also \cite{Libanov:1994ug,Libanov:1995gh,Son:1995wz}). In particular, the computation of correlation functions for charge $n$ operators groups itself in such a way that $n^{-1}$ plays the role of $\hbar$ with $\lambda=g\,n$ acting as the coupling, which naturally suggests to consider the large $n$ limit at fixed $\lambda$. From the point of view of the standard expansion in Feynman diagrams, this limit selects a particular class of diagrams among the many contribution to correlation functions and thus represents a vast simplification.

Inspired by the large charge philosophy, in this work we export those techniques to the analysis of DCFT's. By considering a particular scaling of the defect and bulk couplings we can study in a simple way aspects of the DCFT. In particular, using these techniques, we compute the beta function for the defect couplings and identify the relevant fixed point. We find perfect agreement with the existent literature for line defects in the WF theory near $d=4$ (\textit{e.g.} \cite{Allais:2014fqa,Cuomo:2021kfm}). In turn, the analysis of surface defects in the cubic theory near $d=6$ is, to the best of our knowledge, new. Moreover we compute the profile of the defect recovering the expected functional dependence in both cases. These techniques allow to easily compute correlation functions for configurations of multiple defects. We illustrate this with the simplest case of two parallel defects in both cases. We then combine this approach with more standard large charge techniques to compute correlation functions of a class of defect operators with large charge under the surviving global symmetry after including the defect in both cases.

The limit that we are considering requires us to scale both the defect coupling $h$ and the bulk coupling(s) --let us generically denoted these by $g$-- so that $\lambda=gn^a$, $\nu=h n^{-a}$ ($a=1$ near $d=4$, $a=\frac{1}{2}$ near $d=6$) are kept fixed when $n\rightarrow \infty$. It then turns out that the defect coupling beta function --which is computed exactly in $\nu$ and perturbative only in $\lambda$-- has a zero at $\nu^{2a}\sim\lambda^{-1}\,\epsilon$. This result holds irrespective of having the bulk theory at the fixed point. Tuning it to the fixed point sets $g^{\frac{1}{a}}\sim \epsilon$, which implies that $\nu\sim n^{-\frac{1}{2}}\, \epsilon^{\frac{1-a}{2a}}$. Even though in this limit the defect contribution is suppressed in $n$, it cannot be neglected as it is its leading effect.

The scaling limit presented in this paper can be regarded as a tool to study aspects of the DCFT. In particular, it allows to easily compute the correlator of parallel defects separated in transverse space, as well as correlators of defect operators when combined with large charge techniques (even though the integrals involved are hard to do in general and we concentrated on operators inserted on the defect, \textit{i.e.} defect operators). However, it acquires further interest under the light of \cite{Cuomo:2022xgw} (see also \cite{Beccaria:2022bcr} for a similar idea in the context of Wilson lines in gauge theories), which appeared as this note was being prepared. In that reference it is argued, in the $O(3$) model in $d<4$, that the pinning field defect as studied in this paper is the effective description of an impurity with large spin. In more mundane terms, this means that an atom with spin $n$ acting as an impurity in a antiferromagnetic material is effectively described as an external localized magnetic for large $n$. This gives a further motivation to consider the limit presented in this note, as it has a direct physical application.\footnote{The defect in the cubic model studied in this paper seems to be on a slightly different footing, as it does not break the bulk global symmetry. It would be interesting to study whether it also emerges as an effective description as well.}

This work leaves a number of avenues to explore in the future. To begin with, it would be interesting to extend our results for correlators of large charge operators to generic insertion points and study their implications through the generic restrictions imposed by conformal invariance in the presence of boundaries/defects in \textit{e.g.} \cite{McAvity:1995zd,Billo:2016cpy}. It would also be interesting to study more generic correlators of defects, in particular in more generic arrangements with defects at angles. One could also wonder whether a similar limit can be defined in gauge theories for Wilson lines in higher representations (for instance, the $n$-fold symmetric of the fundamental) by taking large $n$ at fixed $g_{YM}^2n$ --somewhat similarly to \cite{Bourget:2018obm}--, and whether such limit would be related in any way to  \cite{Cuomo:2022xgw,Beccaria:2022bcr}. Another interesting aspect to study is the potential relation to the convexity conjecture in \cite{Aharony:2021mpc} (see also \cite{Antipin:2021rsh} for a discussion in a context related to ours) when applied to defects. We leave these aspects for future studies.

The rest of this paper is organized as follows. In section \ref{d=4} we study line defects in the $O(2N+1)$ Wilson-Fisher theory in $d=4-\epsilon$ in the scaling limit. These defects break the symmetry down to $O(2N)$. From the defect partition function we extract the defect beta function and the location of the defect fixed point. As a by-product we compute the profile of the defect recovering the expected functional form. We then compute correlation functions for operators in the $n$-fold symmetric representation of the surviving $O(2N)$ for large $n$, as well as correlation functions for parallel defects. In section \ref{d=6} we turn to surface defects in the $O(2N)$ theory near $d=6$ using the cubic model in \cite{Fei:2014yja}. The defects that we consider preserve the $O(2N)$ global symmetry. In the scaling limit, we compute the defect beta function and identify a fixed point. The profile of the defect recovers also in this case the expected functional form. We then compute correlation functions of $n$-fold symmetric operators for large $n$ as well as correlation functions for defects themselves. In order not to clutter the presentation, we relegate to the appendices \ref{OSAction} and \ref{A2}-\ref{A7} the details of the computations. Moreover, \ref{A1} contains a summary of the relevant Fourier-transformation formulae.

\section{Line defects near $d=4$ in a scaling limit}
\label{d=4}

The starting point is the theory described by (we work in the euclidean and use conventions as in \cite{Arias-Tamargo:2020fow})

\begin{equation}
\label{quartictheory}
S=\int \frac{1}{2}|\partial\vec{\varphi}|^2+\frac{g}{4}(\vec{\varphi}^2)^2\,,
\end{equation}
where $\vec{\varphi}$ is an $O(2N+1)$ vector. This theory has an IR fixed point in $d=4-\epsilon$ at

\begin{equation}
\label{WF4d}
g_{\star}=\frac{8\pi^2}{2N+9}\epsilon+\mathcal{O}(\epsilon^2)\,.
\end{equation}

We now consider trivial line defect along the coordinate $x^1$ and deform by including the symmetry-breaking line operator in the bulk theory path integral

\begin{equation}
\mathcal{D}(\vec{z})=e^{-h\int d\tau\,\varphi^{2N+1}(\tau,\vec{z})}=e^{-h\int dx\,\varphi^{2N+1}\,\delta_T(\vec{x}-\vec{z})}\,,
\end{equation}
where $\delta_T(\vec{x}-\vec{z})$ stands for the delta function in the transverse space  --which in this case is of dimension $d_T=d-1$ with $d=4-\epsilon$-- supported at $\vec{z}$, being $\vec{z}$ the location of the line defect in the transverse space. In the nomenclature of \cite{Cuomo:2022xgw}, this is the pinning field defect. Note that this line defect breaks the $O(2N+1)$ global symmetry down to $O(2N)$. 

The partition function in the presence of $\mathcal{D}$ is

\begin{equation}
\langle \mathcal{D}(\vec{z})\rangle=\int e^{-\int \frac{1}{2}|\partial\vec{\varphi}|^2+\frac{g}{4}(\vec{\varphi}^2)^2+h\varphi^{2N+1}\,\delta_T(\vec{x}-\vec{z})}\,.
\end{equation}
The presence of the couplings $g$ and $h$ allows for interesting relative scaling limits. In particular, re-scaling the fields $\vec{\varphi}\rightarrow h\vec{\varphi}$, one has

\begin{equation}
\langle \mathcal{D}(\vec{z})\rangle=\int e^{-h^2\int \frac{1}{2}|\partial\vec{\varphi}|^2+\frac{gh^2}{4}(\vec{\varphi}^2)^2+\varphi^{2N+1}\,\delta_T(\vec{x}-\vec{z})}\,.
\end{equation}
This motivates to consider the limit in which $h\rightarrow \infty$ whith $gh^2$ fixed. This results in a new semiclassical approximation where $h$ acts as $\hbar^{-1}$ and the fixed quantity $gh^2$ acts as a coupling which allows for a perturbative treatment of the quartic interaction.\footnote{Note that in this limit the defect comes with strength one. Nevertheless, this presents no technical problem, since, as we will see below, the semiclassical equations of motion can be exactly solved in the defect. One may imagine an alternative limit, where one re-scales $\vec{\varphi}=g^{-\frac{1}{2}}\vec{\varphi}$ and keeps fixed $h\sqrt{g}$. However in this limit the leftover coupling controls the defect insertion, and one would need to exactly take into account the quartic interaction.} To systematize the expansion it is useful to introduce a parameter $n$ and write $\varphi=\sqrt{n}\phi$, $g=\frac{\lambda}{n}$ and $h=\nu\sqrt{n}$. Then

\begin{equation}
\langle \mathcal{D}(\vec{z})\rangle=\int e^{-nS_{\rm eff}},\qquad S_{\rm eff}=\int \frac{1}{2}|\partial\vec{\phi}|^2+\frac{\lambda}{4}(\vec{\phi}^2)^2+\nu\phi^{2N+1}\,\delta_T(\vec{x}-\vec{z})\,.
\end{equation}
Note that introducing the parameter $n$ allows to make natural contact with \cite{Cuomo:2022xgw}, where it is argued that the pinning defect emerges as a large spin limit of an impurity in the $O(N)$ model.

We now take the triple-scaling limit where $n$ to infinity with fixed $\lambda$ and $\nu$. In this regime we can approximate the partition function by the saddle point value 

\begin{equation}
\langle \mathcal{D}(\vec{z})\rangle=e^{-nS_{\rm eff}}\,,
\end{equation}
where now $S_{\rm eff}$ is to be evaluated on the saddle point, whose identification is our next task. The saddle point equations for $S_{\rm eff}$ are

\begin{eqnarray}
\partial^2\phi^a-\lambda\vec{\phi}^2\,\phi^a-\nu\,\delta_T(\vec{x}-\vec{z})\delta^{a,2N+1}=0\,.
\end{eqnarray}
For $a\ne 2N+1$ the solution is $\phi^a=0$. In turn, for $a=2N+1$ we need to solve

\begin{eqnarray}
\partial^2\phi^{2N+1}-\lambda(\phi^{2N+1})^3-\nu\,\delta_T(\vec{x}-\vec{z})=0\,.
\end{eqnarray}
Assuming $\lambda\ll 1$, we can solve this equation in perturbation theory.\footnote{As discussed above, really the expansion parameter is $gh^2=\lambda\nu^2$. Thus, to be fully precise we should demand $\lambda\nu^2\ll 1$.} To first order

\begin{eqnarray}
\partial^2\phi^{2N+1}-\nu\,\delta_T(\vec{x}-\vec{z})=0\,,
\end{eqnarray}
whose solution is

\begin{equation}
\phi^{2N+1}=-\nu\int dy\,G(x-y)\delta_T(\vec{y}-\vec{z})\,.
\end{equation}

We now need to evaluate the action on-shell. Using the equations of motion $S_{\rm eff}$ can be massaged into (see appendix \ref{OSAction} for further details)

\begin{equation}
\label{d=4SIntegral}
S_{\rm eff}=\frac{\nu}{2}\int \phi^{2N+1}\,\delta_T(\vec{x}-\vec{z})+\frac{\lambda}{4}\int (\phi^{2N+1})^4\,.
\end{equation}
The integrals can be done (see appendix \ref{A2} for details of the computation), finding

\begin{eqnarray}
\label{d=4S}
S_{\rm eff}&=&\Big(-\frac{\nu^2}{2}+\frac{\lambda\nu^4}{128\pi^2\epsilon}+\frac{\lambda\nu^4}{128\pi^2}(3-\gamma_E+\log(4\pi)) \Big)\,\int dx^0\int \frac{d^{d-1}\vec{p}}{(2\pi)^{d-1}}\,\frac{1}{\vec{p}^2}\nonumber \\ && -\frac{\lambda\nu^4}{128\pi^2} \int dx^0\int \frac{d^{d-1}\vec{p}}{(2\pi)^{d-1}}\,\frac{\log|p|^2}{\vec{p}^2}\,.
\end{eqnarray}
Regulating the time integral as $\int dx^0=T$, we may write

\begin{eqnarray}
\label{d=4S}
S_{\rm eff}&=&\Big(-\frac{\nu^2}{2}+\frac{\lambda\nu^4}{128\pi^2\epsilon}+\frac{\lambda\nu^4}{128\pi^2}(3-\gamma_E+\log(4\pi)) \Big)\,T\,\int \frac{d^{d-1}\vec{p}}{(2\pi)^{d-1}}\,\frac{1}{\vec{p}^2}\nonumber \\ && -\frac{\lambda\nu^4}{128\pi^2}\,T\,\int \frac{d^{d-1}\vec{p}}{(2\pi)^{d-1}}\,\frac{\log|p|^2}{\vec{p}^2}\,.
\end{eqnarray}

Let us now agree to use Minimal Substraction (MS) to renormalize the divergences of $S_{\rm eff}$. We then define a renormalized coupling

\begin{equation}
\label{nuRenormalized}
\nu=\nu_R+\frac{\lambda \nu_R^3}{32\pi^2\epsilon}\,,
\end{equation}
Re-storing now the factors of the renormalization scale

\begin{equation}
\nu=\mu^{\frac{\epsilon}{2}}\Big(\nu_R+\frac{\lambda \nu_R^3}{2(4\pi)^2\epsilon}\Big)\,,
\end{equation}
Note that this agrees with \cite{Cuomo:2021kfm} upon taking into account the different conventions. 

Since $\nu$ cannot depend on the arbitrary scale $\mu$, it must be that $\mu\frac{d\nu}{d\mu}=0$. Given that, to this order, $\lambda=\lambda_R$ and $\beta_{\lambda}=-\epsilon\lambda_R$ this leads to

\begin{equation}
\mu\frac{d\nu_R}{d\mu}=-\frac{\epsilon}{2}\nu_R+\frac{\lambda\nu_R^3}{(4\pi)^2}\,.
\end{equation}
This beta function has a zero at

\begin{equation}
\label{nuR}
\nu_R^2=\frac{8\pi^2}{\lambda}\epsilon\,,
\end{equation}
which corresponds to the fixed point of the defect theory. 

The defect fixed point given by \eqref{nuR} holds irrespective of whether the bulk theory is tuned to its fixed point. If we do so, using \eqref{WF4d}, one has

\begin{equation}
\nu_R^2=\frac{2N+9}{n}\,.
\end{equation}
Thus, the leading contribution of the defect gives rise to terms subleading in $n$. Note that the computation is exact in $\nu$, and thus it is valid as long as $\lambda$ provides a good loop counting parameter.

\subsection{The defect profile}

\begin{figure}[h!]
\centering
\includegraphics[scale=.25]{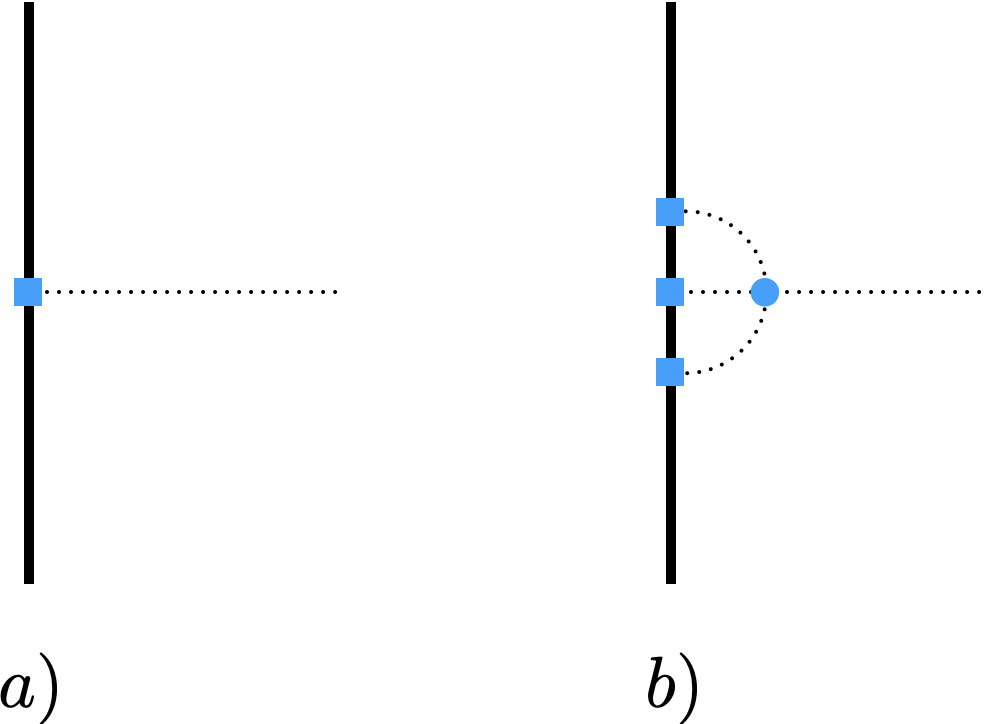}
\caption{Diagramatic expansion of \eqref{Phi2N+1}. We denote by a square the vertex associated to the defect --represented as a thick line-- emitting a $\phi^{2N+1}$ field --represented by a dotted line-- and by a dot the bulk interaction vertex.}
\label{d=4Profile}
\end{figure}

From the definition of $\langle \mathcal{D}\rangle$ it is clear that

\begin{equation}
\label{VEV}
\frac{1}{\langle \mathcal{D}\rangle}\frac{d}{d\nu}\langle \mathcal{D}\rangle=-n \int \langle\phi^{2N+1}\rangle\,\delta_T(\vec{x}-\vec{z})\,.
\end{equation}
In turn, using \eqref{d=4S} and \eqref{nuRenormalized}, we can explicitly compute the LHS. After some massage, we find

\begin{equation}
\int dx\langle \phi^{2N+1}\rangle\delta_T(\vec{x}-\vec{z})=\nu_R\int dx\int \frac{d^{d-1}\vec{p}}{(2\pi)^{d-1}} e^{-i\vec{p}\cdot\vec{x}}\,\Big\{-\frac{1}{\vec{p}^2}+\frac{\lambda\nu_R^2}{32\pi^2}\frac{3-\gamma_E+\log(4\pi)-\log|p|^2}{\vec{p}^2}\Big\}\,\delta_T(\vec{x}-\vec{z})\,,
\end{equation}
Of course, the $\log|\vec{p}|^2$ term has a scale, in which we can re-absorb the finite terms and write

\begin{equation}
\label{Phi2N+1}
\langle \phi^{2N+1}\rangle=\nu_R\int \frac{d^{d-1}\vec{p}}{(2\pi)^{d-1}} e^{-i\vec{p}\cdot\vec{x}}\,\Big\{-\frac{1}{\vec{p}^2}-\frac{\lambda\nu_R^2}{32\pi^2}\frac{\log|p|^2}{\vec{p}^2}\Big\}\,,
\end{equation}
Diagramatically this corresponds to fig.\eqref{d=4Profile}.

Eq.\eqref{Phi2N+1} is the expansion of

\begin{equation}
\langle \phi^{2N+1}\rangle=-\nu_R\int \frac{d^{d-1}\vec{p}}{(2\pi)^{d-1}} \frac{e^{-i\vec{p}\cdot\vec{x}}}{|\vec{p}|^{2-\frac{\lambda\nu_R^2}{16\pi^2}}}\,.
\end{equation}
At the defect fixed point we recover the expected scaling, in agrremeent with eq. (21) in \cite{Allais:2014fqa}

\begin{equation}
\langle \phi^{2N+1}\rangle=-\nu_R\int \frac{d^{d-1}\vec{p}}{(2\pi)^{d-1}} \frac{e^{-i\vec{p}\cdot\vec{x}}}{|\vec{p}|^{2-\frac{\epsilon}{2}}}\sim \frac{1}{|\vec{x}_T|^{\frac{d-2}{2}}}\,,
\end{equation}
where $\vec{x}_T$ refers to location in the transverse space.

\subsection{Correlators in the defect theory}

The defect insertion along $\phi^{2N+1}$ breaks the $O(2N+1)$ symmetry down to $O(2N)$. Introducing complex combinations $\Phi^i$, $i=1\cdots N$ so that only a $U(N)$ is manifest, the action including the defect is

\begin{equation}
S=\int \frac{1}{2}(\partial\varphi^{2N+1})^2+|\partial\vec{\Phi}|^2+\frac{g}{4}\Big((\varphi^{2N+1})^2+2|\vec{\Phi}|^2\Big)^2+h\varphi^{2N+1}\delta_T(\vec{x}-\vec{z})\,.
\end{equation}

We can now consider operators in the $[n,0\cdots0]$ of the unbroken $O(2N)$ symmetry. Following the argument in \cite{Arias-Tamargo:2020fow} (see also \cite{Antipin:2020abu}), their correlator is captured by $\langle (\Phi_1(z_1))^n\,(\Phi_1^{\star}(z_2))^n\rangle$. The path integral formula is

\begin{equation}
\langle (\Phi_1(z_1))^n(\Phi_1^{\star}(z_2))^n\rangle=\frac{1}{\langle \mathcal{D}\rangle} \int e^{-S_{\rm eff}}\,,
\end{equation}
where now

\begin{equation}
S_{\rm eff}=\int \frac{1}{2}(\partial\varphi^{2N+1})^2+|\partial\vec{\Phi}|^2+\frac{g}{4}\Big((\varphi^{2N+1})^2+2|\vec{\Phi}|^2\Big)^2+h\varphi^{2N+1}\delta_T(\vec{x}-\vec{z})-n\log\Phi_1 \delta(x-z_1)-n\log\Phi_1^{\star} \delta(x-z_2)\,.
\end{equation}
We now assume the triple-scaling limit above. Upon re-scaling $\varphi^{2N+1}\rightarrow \sqrt{n}\,\phi$, $\Phi_i\rightarrow \sqrt{n}\,\Psi_i$ and introducing $\lambda=gn$, $h=\sqrt{n}\nu$

\begin{equation}
S_{\rm eff}=n\,\int \frac{1}{2}\partial\phi^2+|\partial\vec{\Psi}|^2+\frac{\lambda}{4}\Big(\phi^2+2|\vec{\Psi}|^2\Big)^2+\nu\phi\delta_T(\vec{x}-\vec{z})-\log\Psi_1 \delta(x-z_1)-\log\Psi_1^{\star} \delta(x-z_2)\,.
\end{equation}
For large $n$ with fixed $\lambda,\nu$ we can use the saddle point approximation. The equations of motion, to leading order for small $\lambda$, are

\begin{equation}
\partial^2\phi-\nu\delta(\vec{x}-\vec{z})=0\,,
\end{equation}
and (it is clear that $\Psi_i=0$ for $i\ne 1$)

\begin{equation}
\partial^2\Psi_1+\frac{1}{\Psi_1^{\star}}\delta(x-z_2)=0\,,\qquad \partial^2\Psi_1^{\star}+\frac{1}{\Psi_1}\delta(x-z_1)=0\,.
\end{equation}
The solution to these equations is

\begin{equation}
\phi=-\nu \int dy\, G(x-y)\,\delta_T(\vec{y}-\vec{z})\,,\quad \Psi_1=\frac{G(z-z_2)}{\sqrt{G(z_1-z_2)}}\,, \quad \Psi_1^{\star}=\frac{G(z-z_1)}{\sqrt{G(z_1-z_2)}}\,.
\end{equation}

\begin{figure}[h!]
\centering
\includegraphics[scale=.25]{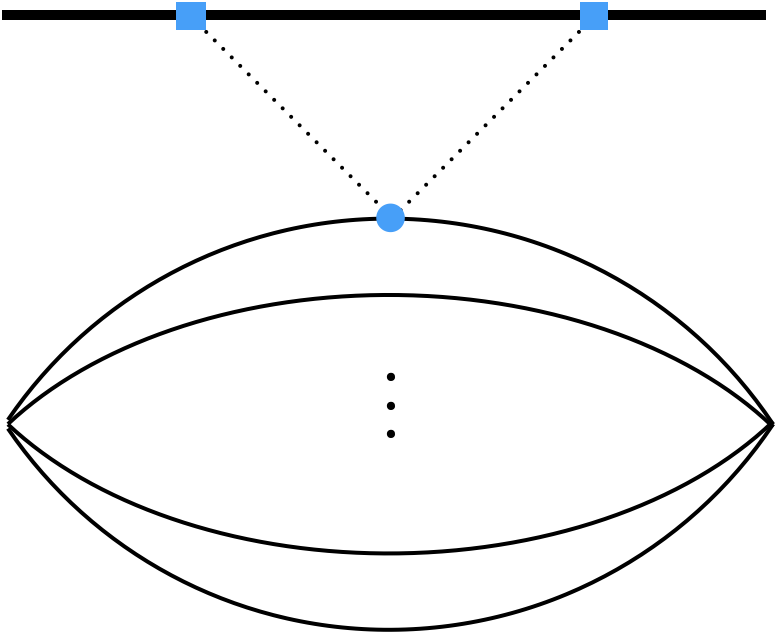}
\caption{Defect correction to the correlation function of defect fields. We denote with solid lines the $\Phi^i$ ($i\ne 2N+1$) fields, whose insertion we put at a generic point for the sake of clarity of the figure.}
\label{d=4DefectFieldCorrelator}
\end{figure}

We now need to evaluate the action on-shell. First, note that

\begin{eqnarray}
S_{\rm eff}&=&n\,\int \frac{1}{2}\partial\phi^2+\frac{\lambda}{4}\phi^4+\nu\phi\delta(\vec{x}-\vec{z})\\ \nonumber && +\int |\partial\vec{\Psi}|^2-\log\Psi_1 \delta(x-z_1)-\log\Psi_1^{\star} \delta(x-z_2)+\lambda\Big(|\vec{\Psi}|^4+|\vec{\Psi}|^2\phi^2\Big)\,.
\end{eqnarray}
The first line gives precisely $\langle \mathcal{D}\rangle$, and therefore

\begin{equation}
\langle (\Phi_1(z_1))^n(\Phi_1^{\star}(z_2))^n\rangle=e^{-n\,\tilde{S}_{\rm eff}}\,,
\end{equation}
where

\begin{equation}
\tilde{S}_{\rm eff}=\int |\partial\vec{\Psi}|^2-\log\Psi_1 \delta(x-z_1)-\log\Psi_1^{\star} \delta(x-z_2)+\lambda\Big(|\vec{\Psi}|^4+|\vec{\Psi}|^2\phi^2\Big)\,.
\end{equation}
Up to a constant, this is

\begin{equation}
\label{d=4CorrelatorIntegral}
\tilde{S}_{\rm eff}=-\log G(z_1-z_2)+\lambda\int |\vec{\Psi}_1|^4+\lambda\int|\vec{\Psi}_1|^2\phi^2\,.
\end{equation}
The first integral is the bulk contribution, just as in \cite{Arias-Tamargo:2019xld}. The second integral, arising from the defect interactions, corresponds to the diagram in fig.\eqref{d=4DefectFieldCorrelator}. It is clear that this diagram is suppressed by an extra $\frac{1}{n}$ with respect to the bulk contribution. This is so because the bulk contribution in large $n$ is dominated by ``kermit" diagrams (see \cite{Arias-Tamargo:2019xld}) which involve merging together 2 lines and so there are $\mathcal{O}(n\,(n-1))$ such diagrams, while the number of those in fig.\eqref{d=4DefectFieldCorrelator} is of order $n$. Nevertheless, even if subleading in $n$, it is however the leading contribution arising from the defect and thus cannot be neglected.

The integral corresponding to the contribution of the defect is very complicated. To make progress, we restrict to correlators of defect operators, that is, we set $\vec{z}_1=\vec{z}_2=\vec{z}$ (and take $\vec{z}=0$ with no loss of generality). In that case, the result is (see appendix \ref{A3} for further details on the computation)

\begin{equation}
\label{d=4Correlator}
\hat{S}_{\rm eff}=2(1+\frac{\lambda}{8\pi^2}+\frac{\lambda\nu^2}{16\pi^2})\,\log|z_1^0-z_2^0|\,.
\end{equation}
Therefore,  putting all together

\begin{equation}
\langle (\Phi_1(z_1))^n(\Phi_1^{\star}(z_2))^n\rangle\sim \frac{1}{|z_1^0-z_2^0|^{2n(1+\frac{\lambda}{8\pi^2}+\frac{\lambda\nu^2}{(4\pi)^2})}}\,.
\end{equation}
It then follows that 

\begin{equation}
\Delta_{[n,0\cdots,0]}=n\,(1+\frac{\lambda}{8\pi^2}+\frac{\lambda\nu^2}{(4\pi)^2})\,.
\end{equation}

Tuning both the defect and the bulk theory to the fixed point, this becomes

\begin{equation}
\Delta_{[n,0\cdots,0]}=n\,(1+\frac{n}{2N+9}\epsilon+\frac{\epsilon}{2})\,.
\end{equation}
Indeed, the defect correction is subleading in $n$ to the bulk contribution. However, to give full meaning to this expression one should relate $\epsilon$ and $n$. One could generically assume $\lambda\sim \epsilon^p$. Since at the fixed point, $\lambda\sim \epsilon n$, this is equivalent to choosing $n\sim \epsilon^{p-1}$, which is consistent for $p\in [0,1)$. Tuning $p$ sufficiently far from 0 one can ensure the self-consistency of the computation above.

\begin{figure}[h!]
\centering
\includegraphics[scale=.2]{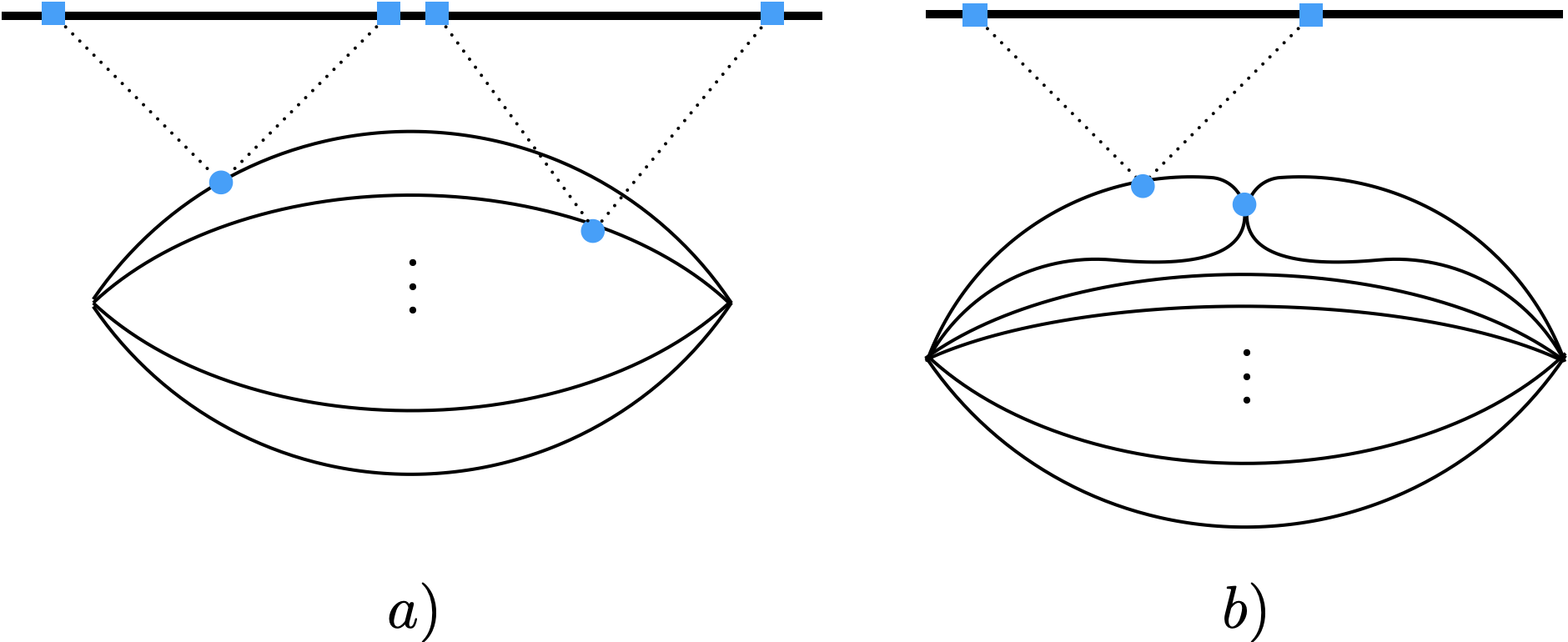}
\caption{Second order corrections to correlators of defect fields.}
\label{d=4SecondOrder}
\end{figure}

It is very interesting however to consider in the detail the second order corrections, where one would find the diagrams in fig.\eqref{d=4SecondOrder}.

It is clear that there are of the order of $n^2$ diagrams of type $a)$, which then all together contribute $n\nu^4\lambda^2\sim n\epsilon^2$, while of type $b)$ there are of the order of $n^2$ diagrams which all together contribute $n^2\nu^2\lambda^2\sim n^3\epsilon^2$. Thus, the $p=0$ case (\textit{i.e.} $\lambda\sim \epsilon^0\ll 1$ would actually select the diagrams of type $b)$ above --which are the generalization of the kermit diagrams in \cite{Arias-Tamargo:2019xld} including the defect. While undoubtedly this case would be very interesting, the relevant integrals are very involved and we have not been able to compute them.

\subsection{Correlators of defects}

Instead of considering a single defect, we could consider the insertion of multiple defects in an arbitrary arrangement. For simplicity, let us consider two parallel defects --that is, oriented along the same direction-- and located, in the transverse space, at $\vec{z}_1$, $\vec{z}_2$ respectively. A quantity of direct interest is the correlation function of two such defects, that is $\langle \mathcal{D}(z_1)\mathcal{D}(z_2)\rangle$. The path integral expression is

\begin{equation}
\langle \mathcal{D}(z_1)\mathcal{D}(z_2)\rangle=\int e^{-\int \frac{1}{2}|\partial\vec{\varphi}|^2+\frac{g}{4}(\vec{\varphi}^2)^2+h\varphi^{2N+1}\,\delta_T(\vec{x}-\vec{z}_1)+h\varphi^{2N+1}\,\delta_T(\vec{x}-\vec{z}_2)}\,.
\end{equation}
Let us now consider the same triple scaling limit as above. Redefining fields and parameters one finds

\begin{equation}
\langle \mathcal{D}(z_1)\mathcal{D}(z_2)\rangle=\int e^{-nS_{\rm eff}},\qquad S_{\rm eff}=\int \frac{1}{2}|\partial\vec{\phi}|^2+\frac{\lambda}{4}(\vec{\phi}^2)^2+\nu\phi^{2N+1}\,\delta_T(\vec{x}-\vec{z}_1)+\nu\phi^{2N+1}\,\delta_T(\vec{x}-\vec{z}_2)\,.
\end{equation}
Taking $n\rightarrow \infty$ for fixed $\lambda$, $\nu$ the integral can be done through saddle point. For small $\lambda$ the saddle point equations set $\phi^a=0$ for $a\ne 2N+1$. For $\phi^{2N+1}$ we have

\begin{eqnarray}
\partial^2\phi^{2N+1}-\nu\,\delta_T(\vec{x}-\vec{z}_1)-\nu\,\delta_T(\vec{x}-\vec{z}_2)=0\,.
\end{eqnarray}
The solution is

\begin{equation}
\phi^{2N+1}=\rho_1(\vec{x})+\rho_2(\vec{x}),\qquad \rho_i(\vec{x})=-\nu\int dy\,G(x-y)\delta_T(\vec{y}-\vec{z}_i)\,.
\end{equation}

Turning to the on-shell action, it can be re-written as\footnote{Note that \textit{a priori} $\int \rho_1^3\rho_2$ and $\int \rho_1\rho_2^3$ seem different. Denoting $i(\vec{z}_1,\vec{z}_2)=\int \rho_1^3\rho_2$, the other integral is $i(\vec{z}_2,\vec{z}_1)=\int \rho_2^3\rho_1$. However by symmetry $i(\vec{z}_1,\vec{z}_2)$ can only involve $|\vec{z}_1-\vec{z}_2|$, so $i(\vec{z}_1,\vec{z}_2)=i(\vec{z}_2,\vec{z}_1)$ (this can be explicitly seen in the result below). Thus these seemingly different contributions are actually equal and thus can be summed up.}

\begin{eqnarray}
S_{\rm eff} &=&\frac{\nu}{2}\int \rho_1\,\delta_T(\vec{x}-\vec{z}_1)+\frac{\lambda}{4}\int \rho_1^4+\frac{\nu}{2}\int \rho_2\,\delta_T(\vec{x}-\vec{z}_2)+\frac{\lambda}{4}\int \rho_1^4\\ \nonumber && +\frac{\nu}{2}\int \rho_1\delta_T(\vec{x}-\vec{z}_2)+\frac{\nu}{2}\int \rho_2\delta_T(\vec{x}-\vec{z}_1)+\frac{\lambda}{2}\int 4\rho_1^3\rho_2+3\rho_1^2\rho_2^2\,.
\end{eqnarray}
We recognize here the contribution of $\mathcal{D}(z_i)$ alone and then an interaction term. Thus

\begin{equation}
\langle \mathcal{D}(z_1)\mathcal{D}(z_2)\rangle=\langle \mathcal{D}(z_1)\rangle \langle \mathcal{D}(z_2)\rangle e^{-nS_{\rm I}}\,,
\end{equation}
with

\begin{equation}
\label{S_eff6d}
S_{\rm I}=\frac{\nu}{2}\int \rho_1\delta_T(\vec{x}-\vec{z}_2)+\frac{\nu}{2}\int \rho_2\delta_T(\vec{x}-\vec{z}_1)+\frac{\lambda}{2}\int 4\rho_1^3\rho_2+3\rho_1^2\rho_2^2\,.
\end{equation}
We recognize here the contribution of the diagrams as in fig.\eqref{d=4DefectCorrelator} below.

\begin{figure}[h!]
\centering
\includegraphics[scale=.25]{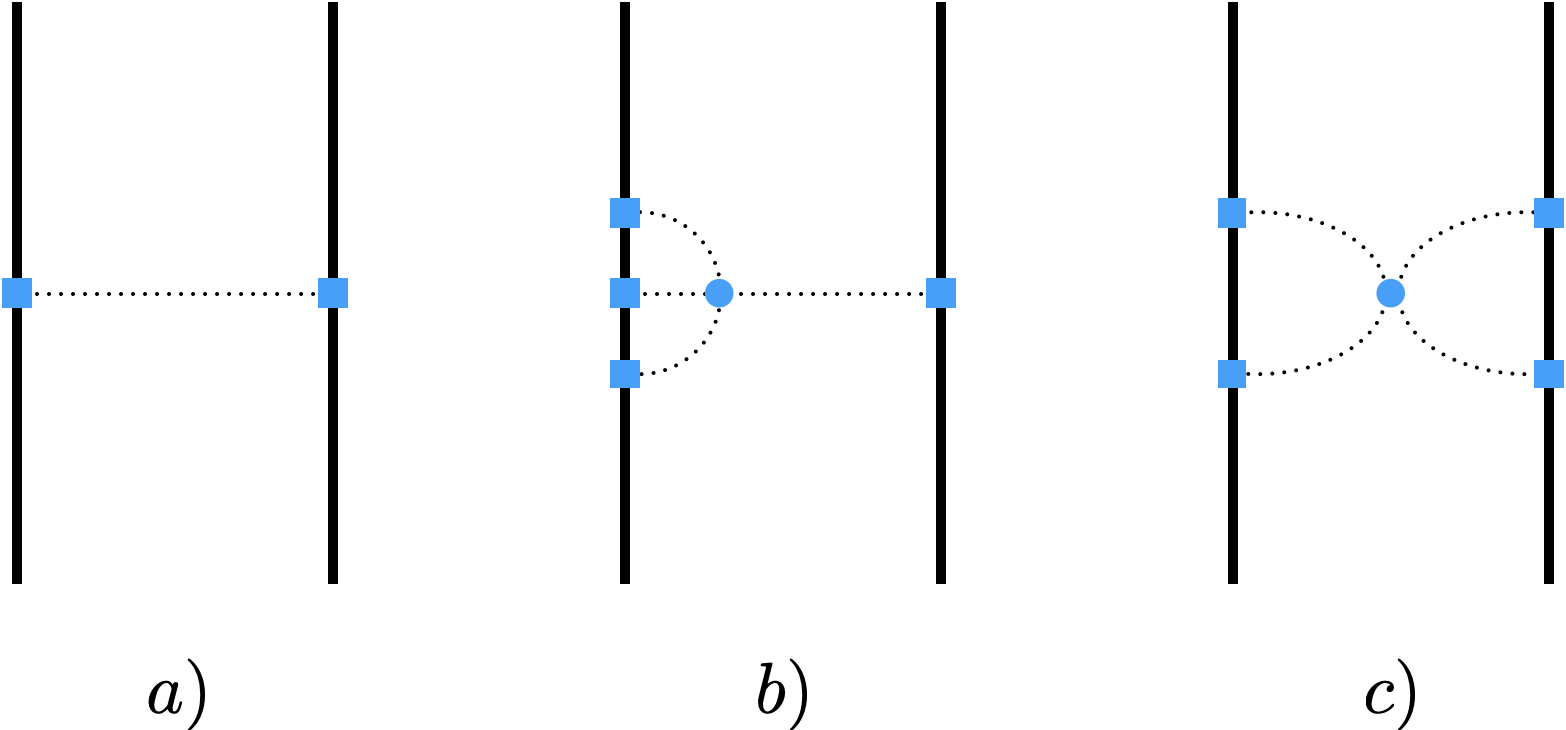}
\caption{Diagrams contributing to the correlation function of line defects: a) corresponds to the $\mathcal{O}(\lambda^0)$ terms; b) corresponds to the $\rho_1^3\rho_2$, $\rho_1\rho_2^3$ contributions; c) corresponds to the $\rho_1^2\rho_2^2$ contribution.}
\label{d=4DefectCorrelator}
\end{figure}

The result of the integrations is (see appendix \ref{A4} for further details)

\begin{eqnarray}
\label{d=4defectcorrelator}
S_{\rm I}&=&\Big[-\frac{\nu^2}{4\pi}\Big(1+\epsilon\log|\vec{z}_1-\vec{z}_2|+\epsilon\frac{\gamma_E+\log(4\pi)}{2}\Big)+\frac{\lambda\nu^4}{64\pi^3\epsilon}+\frac{3\lambda\nu^4}{64\pi^3}\log|\vec{z}_1-\vec{z}_2|\nonumber \\ && -\frac{3\lambda\nu^4}{512\pi}+\frac{3\lambda\nu^4}{128\pi^3}(2+\gamma_E+\log(4\pi))\Big]\,\frac{T}{|\vec{z}_1-\vec{z}_2|}\,.
\end{eqnarray}
Using \eqref{nuRenormalized} to convert the bare couplings in terms of the renormalized couplings, this can be massaged into

\begin{equation}
S_{\rm I}=-\Big(\frac{\nu_R^2}{4\pi}+\frac{3\lambda\nu_R^4}{512\pi}-\frac{\lambda\nu_R^4}{64\pi^3}(3+\gamma_E+\log(4\pi))\Big)\,\frac{T}{|\vec{z}_1-\vec{z}_2|}\,\Big(1-(-\epsilon+\frac{\lambda\nu_R^2}{8\pi^2})\log|\vec{z}_1-\vec{z}_2|\Big)\,.
\end{equation}
We can regard this as the small $\lambda$ expansion of

\begin{equation}
S_{\rm I}=-\Big(\frac{\nu_R^2}{4\pi}+\frac{3\lambda\nu_R^4}{512\pi}-\frac{\lambda\nu_R^4}{64\pi^3}(3+\gamma_E+\log(4\pi))\Big)\,\frac{T}{|\vec{z}_1-\vec{z}_2|^{1+(-\epsilon+\frac{\lambda\nu_R^2}{8\pi^2})}}\,.
\end{equation}
This result agrees, upon turning off the interaction, with \cite{Soderberg:2021kne}. At the defect fixed point the result simplifies, and one finds

\begin{equation}
S_{\rm I}=-\Big(\frac{\nu_R^2}{4\pi}+\frac{3\lambda\nu_R^4}{512\pi}-\frac{\lambda\nu_R^4}{64\pi^3}(3+\gamma_E+\log(4\pi))\Big)\,\frac{T}{|\vec{z}_1-\vec{z}_2|}\,.
\end{equation}

\section{Surface defects near $d=6$ in a scaling limit}
\label{d=6}

Let us consider in $d=6-\epsilon$ the theory with $O(2N)$ global symmetry (we follow conventions of \cite{Arias-Tamargo:2020fow})

\begin{equation}
\label{Scubic}
S=\int \frac{1}{2}|\partial\vec{\varphi}|^2+\frac{1}{2}\partial\eta^2+\frac{g_1}{2}\eta\,|\vec{\varphi}|^2+\frac{g_2}{6}\eta^3\,.
\end{equation}
As shown in \cite{Ma:1975vn} (see also \cite{Fei:2014yja}), this theory admits, in perturbation theory, an IR fixed point for $2N$ larger than a critical $N_{\rm cr}\sim 1038$. The precise location of the fixed point is

\begin{equation}
g_{1\,\star}=\sqrt{\frac{6\,(4\pi)^3}{2N}\epsilon}\,\Big(1+\mathcal{O}(\frac{1}{N})\Big)\,,\qquad g_{2\,\star}=6\sqrt{\frac{6\,(4\pi)^3}{2N}\epsilon}\,\Big(1+\mathcal{O}(\frac{1}{N})\Big)\,.
\end{equation}

The IR fixed point of the cubic theory above has been conjectured to correspond to the completion of the UV fixed point of the quartic theory \eqref{quartictheory} in $d=4+\epsilon$ dimensions \cite{Fei:2014yja}. This is best studied performing a Hubbard-Stratonovich transformation and then tuning to the fixed point. The resulting model is described by an action

\begin{equation}
\label{quarticHS}
S_{\rm quartic}=\int \partial\vec{\xi}^2+\sigma\,\vec{\xi}^2\,,
\end{equation}
where $\vec{\xi}$ is a $O(N)$ vector and $\sigma$ a field with a propagator of the form $\langle \sigma(x)\sigma(x)\rangle\sim |x|^{-4}$. In this guise, the conjectured equivalence between the IR/UV fixed point of the cubic/quartic theory has passed several non-trivial checks, including in the sector of large charge operators as in \cite{Arias-Tamargo:2020fow} (see also \cite{Antipin:2021jiw,Jack:2021ziq} and \cite{Giombi:2020enj}). Nevertheless, the fact that the potential is cubic --and hence not bounded below-- or, alternatively from the quartic model point of view, has negative coupling at the fixed point, ends uf manifesting in the existence of instanton corrections giving imaginary parts to anomalous dimensions \cite{Giombi:2019upv}.

In the cubic theory \eqref{Scubic} we can imagine inserting a trivial surface defect (that is, supported on a 2d space). Let us consider our defect along $(x^4,x^5)$ at $\vec{x}=(x^0,x^1,x^2,x^3)=0$. We will denote with the subscript $_{||}$ the directions parallel to the defect --the worldvolume-- and with $_{T}$ the directions transverse to the defect. In this case $d_T=d-2$ with $d=6-\epsilon$. Then, one may deform the otherwise trivial defect theory with the insertion of 

\begin{equation}
\label{d=6Defect}
\mathcal{D}=e^{-h\int d^2x\,\eta}\,.
\end{equation}
Note that the defect does not break the $O(N)$ symmetry present in the original theory, and is, in that respect, different from the pinning field defect in the nomenclature of \cite{Cuomo:2022xgw}. It is natural to guess that it corresponds, in the quartic theory avatar in eq.\eqref{quarticHS}, to a defect operator  $\mathcal{D}_{\rm quartic}={\rm exp}(-\hat{h}\int d^2x\,\vec{\xi}^2)$.

One could imagine deforming instead with the symmetry-breaking insertion $\mathcal{D}'={\rm exp}(-h'\int d^2x\, \varphi^{2N})$, which is the direct analog of the deformation in $d=4-\epsilon$ (which is, properly speaking, the pinning field defect in the nomenclature of \cite{Cuomo:2022xgw}). However, since there is no potential for $\vec{\varphi}$ alone, this defect would behave as in the free theory (up to higher corrections in $n^{-1}$). For this reason, we restrict to the symmetry-preserving defect operator in eq.\eqref{d=6Defect}.

The defect VEV is

\begin{equation}
\langle \mathcal{D}\rangle=\int e^{-S{\rm eff}}\,,
\end{equation}
with

\begin{equation}
S_{\rm eff}=\int \frac{1}{2}|\partial\vec{\varphi}|^2+\frac{1}{2}\partial\eta^2+\frac{g_1}{2}\eta\,|\vec{\varphi}|^2+\frac{g_2}{6}\eta^3+h\eta\,\delta_T(\vec{x})\,.
\end{equation}
Just as in the 4d case, the couplings $g_i$, $h$ allow for interesting limits when the couplings scale relatively in the appropriate way.  The most interesting such scaling is when $(\vec{\varphi},\eta)\rightarrow h(\vec{\varphi},\eta)$ while keeping $g_ih$ fixed, where a new semiclassical limit with $h^{-1}\sim \hbar$ emerges controlling $g_ih$ the cubic interactions and thus allowing for a systematic perturbative expansion. Just as in the 4d case, it is useful to introduce a parameter $n$ and scale now $\vec{\varphi}=\sqrt{n}\vec{\phi}$, $\eta=\sqrt{n}\rho$ while introducing $g_i\sqrt{n}=h_i$, $h=\nu\sqrt{n}$. Then

\begin{equation}
S_{\rm eff}=n\int \frac{1}{2}|\partial\vec{\phi}|^2+\frac{1}{2}\partial\rho^2+\frac{h_1}{2}\rho\,|\vec{\phi}|^2+\frac{h_2}{6}\rho^3+\nu\rho\,\delta_T(\vec{x})\,.
\end{equation}

In large $n$ with $h_i,\nu$ fixed, we can use the saddle point approximation. Moreover, we will assume $h_1\sim h_2\ll 1$.\footnote{Just as in the 4d case, strictly speaking, the expansion parameter is $h_i\nu\ll 1$.} Then, the saddle point equations set $\vec{\phi}=0$. As for $\rho$, for small $h_i$, we can keep to leading order

\begin{equation}
\partial^2\rho-\nu\delta_T(\vec{x})=0\,.
\end{equation}
The solution is

\begin{equation}
\label{rhoD}
\rho=-\nu \int dz\,G(x-z) \delta_T(\vec{z})\,.
\end{equation}

To evaluate the on-shell action, we need to evaluate, on the saddle point solution (see appendix \ref{OSAction} for further details)

\begin{equation}
\label{d=6SIntegral}
S_{\rm eff}=n\frac{\nu}{2}\int \rho\,\delta_T(x)+ n\, \frac{h_2}{6}\int \rho^3\,.
\end{equation}
The result is (see appendix \ref{A5} for details)

\begin{equation}
\label{d=6S}
S_{\rm eff}=-nV \Big\{\Big(\frac{\nu^2}{2}+ \frac{h_2\nu^3}{48\pi^2}\frac{1}{\epsilon}\Big)\int\frac{d\vec{p}_T}{(2\pi)^{d_T}} \frac{1}{\vec{p}_T^2}-\frac{h_2\nu^3}{48\pi^2}\int\frac{d\vec{p}_T}{(2\pi)^{d_T}}\frac{\log|\vec{p}_T|}{\vec{p}_T^2}\Big\}\,,
\end{equation}
where $\int d\vec{x}_{||}=V$. Using MS to re-absorb divergences, we introduce

\begin{equation}
\nu=\nu_R-\frac{h_2\nu_R^2}{16\pi^2\epsilon}\,.
\end{equation}
Recovering explicitly the dependence on the renormalization scale, this is

\begin{equation}
\nu=\mu^{\frac{\epsilon}{2}}\Big(\nu_R-\frac{h_2\nu_R^2}{16\pi^2\epsilon}\Big)\,.
\end{equation}
Using that $\beta_{h_2}=-\frac{\epsilon}{2}h_2+\cdots$, we find

\begin{equation}
\beta_{\nu}=-\frac{\epsilon}{2}\nu_R-\frac{h_2\nu_R^2}{16\pi^2}\,.
\end{equation}
This has a fixed point at

\begin{equation}
\label{fixedpoint6d}
\nu_R=-\frac{8\pi^2\epsilon}{h_2}\,.
\end{equation}
The presence of the defect fixed point holds irrespective of the bulk theory being at its fixed point. If in addition we tune it to the fixed point

\begin{equation}
\nu_R=-\sqrt{\frac{\pi\,N\,\epsilon}{108\,n}}\,.
\end{equation}
We again see that the contributions of the defect are subleading in $n$. 

As a by-product, we can also compute the dimension of $\hat{\rho}$ following the trick in \cite{Gubser:2008yx,Cuomo:2021kfm}, finding

\begin{equation}
\Delta(\hat{\rho})=2+\frac{\partial \beta}{\partial\nu}\Big|_{\nu_R}=2+\frac{\epsilon}{2}+\mathcal{O}(\epsilon^2)\,.
\end{equation}

\subsection{The defect profile}

\begin{figure}[h!]
\centering
\includegraphics[scale=.25]{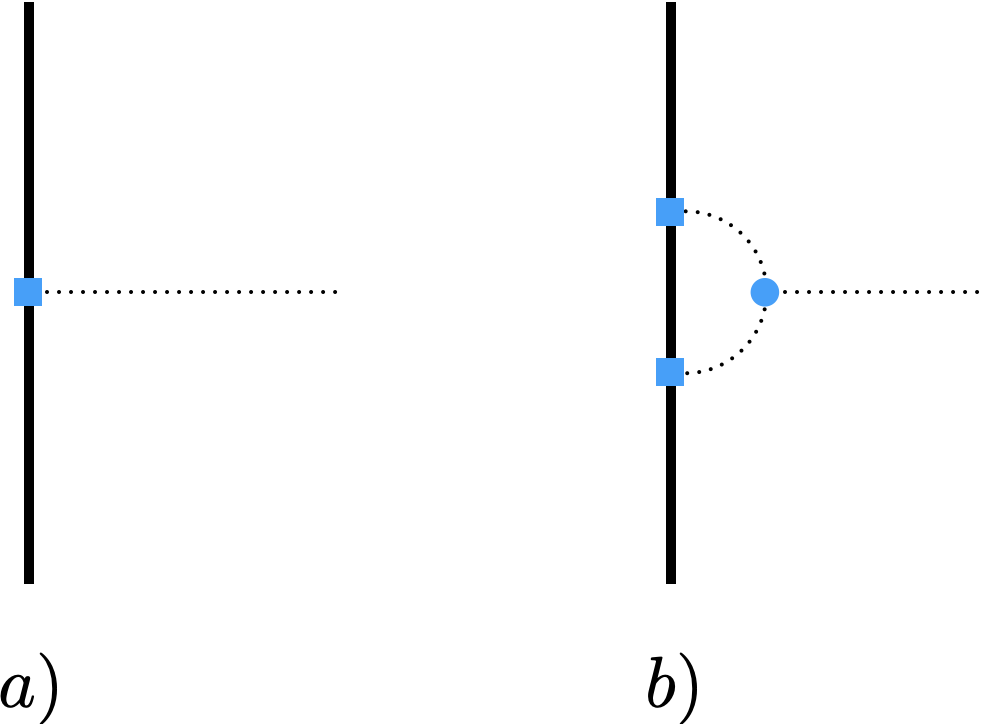}
\caption{Diagramatic expansion of \eqref{rho}. We denote by a square the vertex associated to the defect --represented as a thick line-- emitting a $\rho$ field --represented by a dotted line-- and by a dot the bulk interaction vertex.}
\label{d=6Profile}
\end{figure}

We can compute the profile of the defect following the same strategy as in eq.\eqref{VEV}. To begin with, in terms of the renormalized couplings we have

\begin{equation}
S_{\rm eff}=-nV \Big\{\frac{\nu_R^2}{2}\int\frac{d\vec{p}_T}{(2\pi)^{d_T}} \frac{1}{\vec{p}_T^2}-\frac{h_2\nu_R^3}{48\pi^2}\int\frac{d\vec{p}_T}{(2\pi)^{d_T}}\frac{\log|\vec{p}_T|}{\vec{p}_T^2}\Big\}\,.
\end{equation}
Therefore, using the same trick as in \eqref{VEV}, we now find

\begin{equation}
\label{rho}
\int dx\, \langle \rho\rangle\, \delta_T(\vec{x})=-V\nu_R \int\frac{d\vec{p}_T}{(2\pi)^{d_T}} \Big(\frac{1}{\vec{p}_T^2}-\frac{h_2\nu_R}{16\pi^2}\frac{\log|\vec{p}_T|}{\vec{p}_T^2}\Big)\,.
\end{equation}
Diagramatically this corresponds to fig.\eqref{d=6Profile}.

From eq.\eqref{rho} we can read off 

\begin{equation}
\langle \rho\rangle=- \nu_R \int\frac{d\vec{p}_T}{(2\pi)^{d_T}} \frac{e^{i\vec{p}_T\cdot\vec{x}_T}}{|\vec{p}_T|^{2+\frac{h_2\nu_R}{16\pi^2}}}\,.
\end{equation}
At the fixed point we recover the expected scaling

\begin{equation}
\label{comment}
\langle \rho\rangle=-\nu_R \int\frac{d\vec{p}_T}{(2\pi)^{d_T}} \frac{e^{i\vec{p}_T\cdot\vec{x}_T}}{|\vec{p}_T|^{2-\frac{\epsilon}{2}}}\sim \frac{1}{|\vec{x}_T|^{\frac{d-2}{2}}}\,.
\end{equation}

\subsection{Correlators in the defect theory}

We can consider operators in the $[n,0\cdots0]$ of the unbroken $O(2N)$ symmetry. By the argument in \cite{Arias-Tamargo:2020fow} (see also \cite{Antipin:2020abu}), their correlator, in the presence of the defect is captured by $\langle (\Phi_1(z_1))^n\,(\Phi_1^{\star}(z_2))^n\rangle$, which can be computed as

\begin{equation}
\langle (\Phi_1(z_1))^n\,(\Phi_1^{\star}(z_2))^n\rangle=\frac{1}{\langle \mathcal{D}\rangle}\int e^{-S_{\rm eff}}\,.
\end{equation}
where ($\Phi_1\equiv \Phi$. Moreover, with no loss of generality, we consider the defect at the origin)

\begin{equation}
S_{\rm eff}=n\int \frac{1}{2}|\partial\Phi|^2+\frac{1}{2}\partial\rho^2+\frac{h_1}{2}\rho\,|\Phi|^2+\frac{h_2}{6}\rho^3+\nu\rho\,\delta_T(x)-\log\Phi\delta(x-z_1)-\log\Phi^{\star}\delta(x-z_2)\,.
\end{equation}

At weak coupling the relevant saddle point equations are

\begin{equation}
\partial^2\Phi+\frac{1}{\Phi^{\star}}\delta(x-z_2)=0\,,\qquad \partial^2\Phi^{\star}+\frac{1}{\Phi}\delta(x-z_1)=0\,.
\end{equation}
and

\begin{equation}
\partial^2\rho-\nu\delta(x)-\frac{h_1}{6}|\Phi|^2=0\,.
\end{equation}
The solution to these equations is

\begin{equation}
\Phi=\Phi_C,\qquad \Phi^{\star}=\Phi_C^{\star},\qquad \rho=\rho_D+\rho_C\,,
\end{equation}
with $\Phi_C$, $\Phi_C^{\star}$ and $\rho_C$ the solutions in \cite{Arias-Tamargo:2020fow}. Recall that $\rho_C$ is itself of order $h_1$. Moreover, $\rho_D$ is given by \eqref{rhoD}, which in position space reads 

\begin{equation}
\rho_D=-\frac{\nu}{4\pi^2} \frac{1}{|\vec{x}_T|^2}\,.
\end{equation}

\begin{figure}[h!]
\centering
\includegraphics[scale=.25]{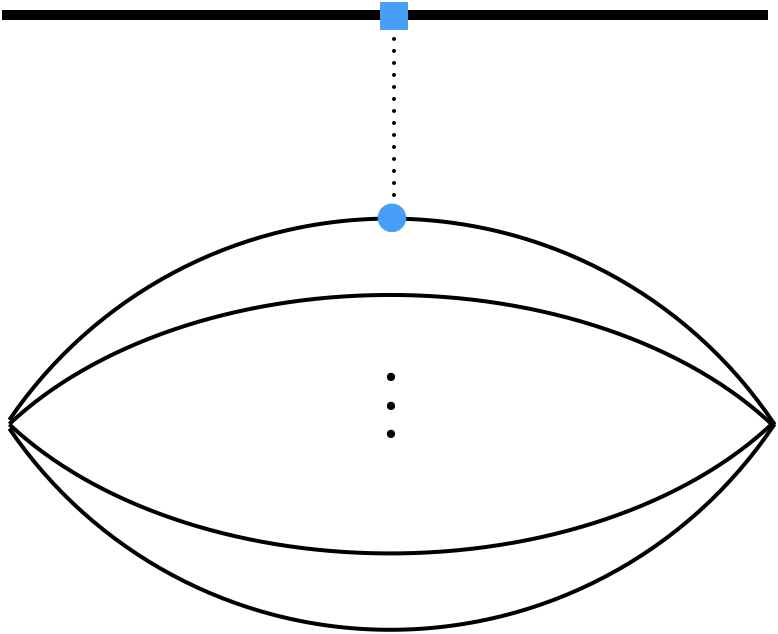}
\caption{Defect correction to the correlation function of defect fields. We denote with solid lines the $\Phi$ fields, whose insertion we put at a generic point for the sake of clarity of the figure.}
\label{d=6DefectFieldCorrelator}
\end{figure}

Evaluating the action one finds, to leading order in $h_i$

\begin{equation}
S_{\rm eff}=S_{\rm eff}^C+S_{\rm eff}^D+\frac{h_1n}{2}\int \rho_D|\Phi_C|^2+n\nu \int \rho_C \delta_T\,,
\end{equation}
where $S_{\rm eff}^{C,D}$ stands for the action evaluated on the $_{C,D}$ fields, respectively. Therefore

\begin{equation}
\label{d=6Correlator}
\langle (\Phi_1(z_1))^n\,(\Phi_1^{\star}(z_2))^n\rangle=e^{-S_{\rm eff}^C-\frac{h_1n}{2}\int \rho_D|\Phi_C|^2-n\nu \int \rho_C \delta_T}\,.
\end{equation}
Diagramatically, the defect contribution to the correlation function is coming from the diagram in fig.\eqref{d=6DefectFieldCorrelator}. Because of the same reason as above, it is clear that this diagram is subleading in $n$ when compared to the bulk contribution. Yet, as it is the leading contribution arising from the presence of the defect it cannot be neglected.

The contribution of the defect is a complicated integral. To make progress, we concentrate on defect operators by taking $\vec{z}_{i,T}=0$. Then, doing the integrals one finds (see appendix \ref{A6} for further details)

\begin{equation}
\label{d=6correlator}
\langle (\Phi_1(z_1))^n\,(\Phi_1^{\star}(z_2))^n\rangle\sim \frac{1}{|\vec{z}_{1||}-\vec{z}_{2||}|^{2n\,(2-\frac{h_1^2}{64\pi^3}-\frac{3h_1\nu}{32\pi^2})}}\,,
\end{equation}
This suggests that

\begin{equation}
\Delta_{[n,0\cdots0]}=n\,\Big(2-\frac{h_1^2}{64\pi^3}-\frac{3h_1\nu}{32\pi^2}\Big)\,.
\end{equation}
Note that at the fixed point $\nu<0$, so the second term is actually positive. Tuning the theory to the bulk and defect fixed points

\begin{equation}
\Delta_{[n,0\cdots0]}=n\,\Big(2-\frac{3\,n}{N}\epsilon+\frac{\epsilon}{8}\Big)\,.
\end{equation}
Note that the same remarks as in the $d=4$ case above apply.

It would be very interesting to study the convexity properties of the DCFT under the light of \cite{Aharony:2021mpc} (see also \cite{Antipin:2021rsh} for a discussion in a context related to ours). The defect interaction contributes oppositely to the bulk piece, but is suppressed so as to be able to compensate for it. It would be interesting to study whether for $N$ sufficiently large there is a regime where the defect contribution can change the convexity properties and the implications of this for the weak gravity conjecture.

\subsection{Correlators of defects}

Let us consider now two parallel defects, located in transverse space one at $\vec{z}_{1T}$ and the other at $\vec{z}_{2T}$. Their correlator is computed as

\begin{equation}
\langle \mathcal{D}(\vec{z}_{1T})\mathcal{D}(\vec{z}_{2T})\rangle=\frac{1}{\langle \mathcal{D}(\vec{z}_{1T})\rangle \langle \mathcal{D}(\vec{z}_{2T})\rangle} \int e^{-S_{\rm eff}}\,,
\end{equation}
with

\begin{equation}
S_{\rm eff}=n\int \frac{1}{2}|\partial\vec{\phi}|^2+\frac{1}{2}\partial\rho^2+\frac{h_1}{2}\rho\,|\vec{\phi}|^2+\frac{h_2}{6}\rho^3+\nu\rho\,\delta_T(\vec{z}_{1T})+\nu\rho\,\delta_T(\vec{z}_{2T})\,.
\end{equation}
To leading order in the $h$'s, the saddle point solution is

\begin{equation}
\rho=\rho_1+\rho_2,\qquad \rho_i=-\nu\int \frac{d^{d_T}\vec{p}_T}{(2\pi)^{d_T}}\,\frac{e^{i\vec{p}_T\cdot(\vec{x}_T-\vec{z}_{iT})}}{\vec{p}_T^2}\,.
\end{equation}
Evaluating the action on-shell gives 

\begin{equation}
S_{\rm eff}=nS_{\rm eff}^{(1)}+nS_{\rm eff}^{(2)}+n \int \frac{3h_2}{6}\rho_1^2\rho_2+ \frac{3h_2}{6}\rho_2^2\rho_1+\frac{1}{2}\nu\rho_2\,\delta_T(\vec{z}_{1T})+\frac{1}{2} \nu\rho_1\,\delta_T(\vec{z}_{2T})\,,
\end{equation}
where $S_{\rm eff}^{(i)}$ is the on-shell action for defect $i$. Hence

\begin{equation}
\label{d=6DefectCorrelatorIntegral}
\langle \mathcal{D}(\vec{z}_{1T})\mathcal{D}(\vec{z}_{2T})\rangle=e^{-nS_{\rm eff}}\,,\qquad S_{\rm eff}=\int h_2\rho_1^2\rho_2+\nu\rho_2\,\delta_T(\vec{z}_{1T})\,.
\end{equation}
We recognize here the contribution of the diagrams as in fig.\eqref{d=6DefectCorrelator} below.

\begin{figure}[h!]
\centering
\includegraphics[scale=.25]{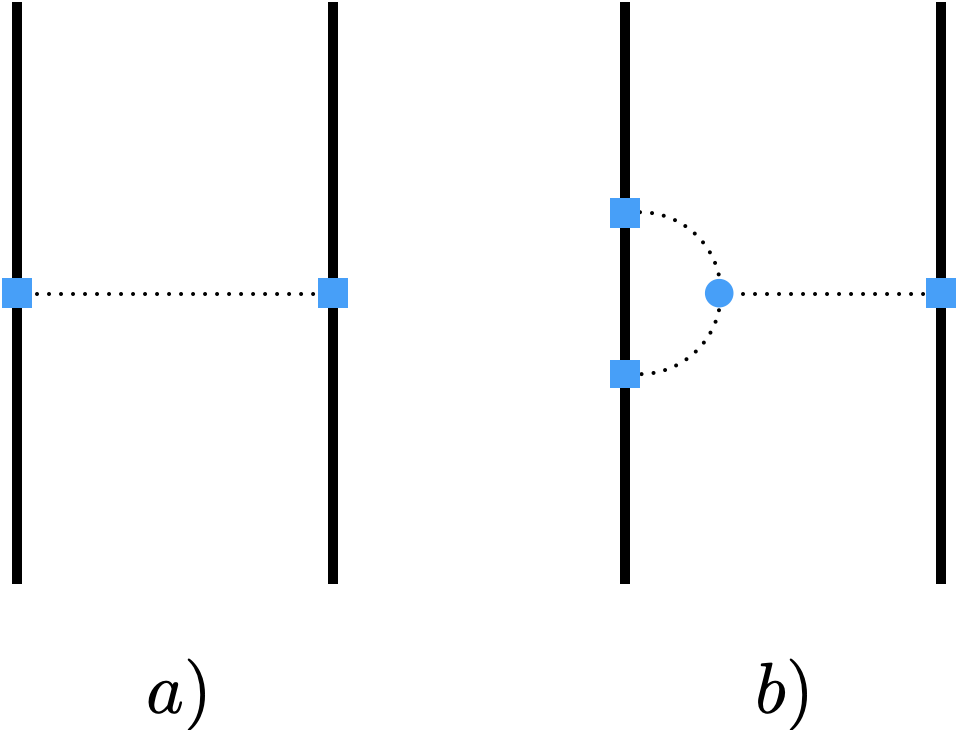}
\caption{Diagrams contributing to the correlation function of line defects: a) corresponds to the $\mathcal{O}(h_i^0)$ terms; b) corresponds to the $\rho_1^2\rho_2$ contribution.}
\label{d=6DefectCorrelator}
\end{figure}

Doing the integrals, one finds (see appendix \ref{A7} for details)

\begin{equation}
\label{d=6DefectCorrelatorResult}
S_{\rm eff}= -\Big(\frac{\nu_R^2}{4\pi^2}+\frac{\nu_R^3h_2}{64\pi^4}\,(2+\gamma_E+\log\pi)\Big) \frac{V}{|\vec{z}_{1T}-\vec{z}_{2T}|^{2-(\epsilon+\frac{h_2\nu_R}{8\pi^2})}}\,.
\end{equation}
Upon putting the defect theory at the fixed point using \eqref{fixedpoint6d}, one finds

\begin{equation}
\label{d=6DefectCorrelatorResult}
S_{\rm eff}= -\Big(\frac{\nu_R^2}{4\pi^2}+\frac{\nu_R^3h_2}{64\pi^4}\,(2+\gamma_E+\log\pi)\Big) \frac{V}{|\vec{z}_{1T}-\vec{z}_{2T}|^{2}}\,.
\end{equation}
It is interesting to note that in this computation the diagrams corresponding to the anomalous dimension of the scalars are suppressed. \footnote{Similar comments apply for the computation of the defect profile in eq.\eqref{comment}. Also the discussion could be extended to the 4d case in the previous section, although in that case the anomalous dimension of the exchanged field enters at 2 loops.} Indeed, going to the next order in perturbation theory one would have 3 diagrams: one which is two copies of that in fig.7.b); one which is a ``ladder" (two copies like fig.7.a) connected by an intermediate line) and the one-loop self-energy of the exchanged field in fig.7.a). To understand those, it is perhaps easiest to use the Feynman rules from the original lagrangian. The first two diagrams would contribute $\sim h^4g_2^2\sim n\,h_2^2\nu^2$ while the last would contribute $\sim h^2g_2^2\sim h_2^2\nu^2$. As a consequence, the self-energy diagram would actually be suppressed. This can also be understood ``combinatorially": in terms of $m\sim \sqrt{n}$, the defect is akin to $m$ defects of coupling $\nu$. Then, hanging 2 lines on one such defect can be done in $\sim m^2$ ways, while hanging only one line can be done in $m$ ways. As a consequence, the one-loop diagram is to be expected to be suppressed by a factor of $m^2=n$ with respect to the other 2 diagrams. In this respect, our limit is in fact selecting a subset of diagrams. Note however that the fixed point is found at small values of the double-scaling parameter. Thus, it would be interesting to further study higher orders.

\section*{Akcnowledgements}

I would like to thank Jorge Russo for many enlightening conversations. This work is partly supported by Spanish national grant MINECO-16-FPA2015- 63667-P as well as the Principado de Asturias grant SV-PA-21-AYUD/2021/52177.  

\begin{appendix}

\section{On the evaluation of the on-shell action}\label{OSAction}

In this appendix we provide further details on the computation of the on-shell action. To cover both the $d=4,\,6$ cases, consider the action

\begin{equation}
S=\int dx\,\frac{1}{2}\partial\phi^2+\frac{\lambda}{nq}\phi^n+\nu\,\phi\,\delta(\vec{x}-\vec{z})\,.
\end{equation}
Upon suitably choosing the parameters $n,\lambda,q,\nu$ here, this is the relevant part of the action for both cases. The equation of motion is

\begin{equation}
\partial^2\phi-\frac{\lambda}{q}\,\phi^{n-1}-\nu\,\delta(\vec{x}-\vec{z})=0\,.
\end{equation}
Let us solve this equation in perturbation theory to order $\lambda$. Writting $\phi=\phi_0+\lambda\,\phi_1$, one has

\begin{equation}
\partial^2\phi_0=\nu\,\delta(\vec{x}-\vec{z})\,,\qquad \partial^2\phi_1=\frac{1}{q}\phi_0^{n-1}\,.
\end{equation}
The solution for $\phi_0$ is

\begin{equation}
\phi_0=\nu \int dy\,G(x-y)\,\delta(\vec{x}-\vec{y})\,.
\end{equation}
In addition, the solution for $\phi_1$ is

\begin{equation}
\phi_1=\frac{1}{q}\int dy G(x-y)\,\phi_0^{n-1}(y)\,.
\end{equation}

Let us now go back to $S$. Using the full eom. one has

\begin{equation}
S_{\rm os}=\int dx \nu\,\phi\,\delta(\vec{x}-\vec{z})-\int dx \lambda \frac{n-2}{2nq}\phi^n\,.
\end{equation}
Plugging here the solution to order $\lambda$

\begin{equation}
S_{\rm os}=\frac{\nu}{2}\,\int dx\,\phi_0\,\delta(\vec{x}-\vec{z})+\frac{\lambda\nu}{2}\, \int dx \phi_1\,\delta(\vec{x}-\vec{z})-\int dx \lambda \frac{n-2}{2nq}\phi_0^n\,.
\end{equation}
Now

\begin{equation}
\int dx \phi_1\,\delta(\vec{x}-\vec{z})=\frac{1}{q} \int dx \int dy G(x-y)\,\phi_0^{n-1}(y)\,\delta(\vec{x}-\vec{z})=  \frac{1}{q}\int dy\,\phi_0^{n-1}(y)\, \int dx G(x-y)\,\delta(\vec{x}-\vec{z})\,.
\end{equation}
Up to a $\nu^{-1}$, we recognize here $\phi_0$, so

\begin{equation}
\int dx \phi_1\,\delta(\vec{x}-\vec{z})= \frac{1}{\nu q} \int dy\,\phi_0^n(y)\,.
\end{equation}
So

\begin{equation}
S_{\rm os}=\frac{\nu}{2}\,\int dx\,\phi_0\,\delta(\vec{x}-\vec{z})+\frac{\lambda}{2q}\, \int dx \phi_1^n-\int dx \lambda \frac{n-2}{2nq}\phi^n=\frac{\nu}{2}\,\int dx\,\phi_0\,\delta(\vec{x}-\vec{z})+\frac{\lambda}{nq}\, \int dx \phi_1^n\,.
\end{equation}
Thus, all in all we find the on-shell action as described in the main text.

\section{Further details on the integrals}

\subsection{Fourier transform formulae}
\label{A1}

For Fourier transforms we use

\begin{equation}
\frac{1}{(x^2)^{\alpha}}=\frac{(4\pi)^{\frac{d}{2}}\,\Gamma(\frac{d}{2}-\alpha)}{4^{\alpha}\Gamma(\alpha)}\,\int \frac{d^dp}{(2\pi)^d}\,\frac{e^{-ipx}}{(p^2)^{\frac{d}{2}-\alpha}}\,.
\end{equation}

In particular one has

\begin{equation}
G(x)=\int \frac{d^dp}{(2\pi)^d}\,\frac{e^{-ipx}}{p^2}=\frac{\Gamma(\frac{d}{2}-1)}{4\pi^{\frac{d}{2}}}\,\frac{1}{(x^2)^{\frac{d-2}{2}}}\,.
\end{equation}

\subsection{Eq.\eqref{d=4S}}
\label{A2}

In this appendix we compile details of the derivation of \eqref{d=4S} starting with \eqref{d=4SIntegral}. Fourier-transforming the first term in  \eqref{d=4SIntegral} gives

\begin{equation}
\frac{\nu}{2}\int \phi^{2N+1}\,\delta(\vec{x}-\vec{z})= -\frac{\nu^2}{2} \int dx^0 \int\frac{d^{d-1}\vec{p}}{(2\pi)^{d-1}}\,\frac{1}{\vec{p}^2}\,.
\end{equation}

As for the second term, upon Fourier transforming and appropriately shifting the integration variables, we find 

\begin{equation}
\frac{\lambda}{4}\int (\phi^{2N+1})^4=\frac{\lambda\nu^4}{4}\int dx^0\int \frac{d^{d-1}\vec{p}_1}{(2\pi)^{d-1}}\,\frac{1}{\vec{p}_1^2}\,\int\frac{d^{d-1}\vec{p}_2}{(2\pi)^{d-1}}\,\int\frac{d^{d-1}\vec{p}_3}{(2\pi)^{d-1}}\, \,\frac{1}{\vec{p}_2^2\,(\vec{p}_3+\vec{p}_1)^2\,(\vec{p}_2-\vec{p}_3)^2}
\end{equation}
Using eq. 2.8 in \cite{Grozin:2003ak} with $n_1=0$, $n_2=1$, $n_3=1$, $n_4=0$ and $n_5=1$

\begin{equation}
\frac{\lambda}{4}\int (\phi^{2N+1})^4=\frac{\pi^{d-1}}{(2\pi)^{2d-2}}G(0,1,1,0,1) \frac{\lambda\nu^4}{4}\int dx^0\int \frac{d^{d-1}\vec{p}_1}{(2\pi)^{d-1}}\,\frac{1}{(\vec{p}_1^2)^{5-d}}\,.
\end{equation}
Using (2.10) in that reference, in $d=4-\epsilon$ dimensions one finds

\begin{equation}
G(0,1,1,0,1)=G(1,1)\,G(1,2-\frac{d-1}{2})=\frac{2\pi}{\epsilon}+(3-\gamma_E)2\pi\,.
\end{equation}
Therefore

\begin{equation}
\frac{\lambda}{4}\int (\phi^{2N+1})^4= \frac{\lambda\nu^4}{128\pi^2}\,\Big(\frac{1}{\epsilon}+3-\gamma_E+\log(4\pi)) \Big) \, \int dx^0\int \frac{d^{d-1}\vec{p}}{(2\pi)^{d-1}}\,\frac{1}{\vec{p}^2}-\frac{\lambda\nu^4}{128\pi^2} \int dx^0\int \frac{d^{d-1}\vec{p}}{(2\pi)^{d-1}}\,\frac{\log|p|^2}{\vec{p}^2}\,.
\end{equation}

\subsection{Eq.\eqref{d=4Correlator}}
\label{A3}

In this appendix we describe the derivation of \eqref{d=4Correlator}.  The starting point is \eqref{d=4CorrelatorIntegral}. The first integral corresponds to the bulk contribution, and can be borrowed from \cite{Arias-Tamargo:2019xld}

\begin{equation}
\lambda\int |\vec{\Psi}_1|^4=\frac{\lambda}{4\pi^2}\log|z_1-z_2|\,.
\end{equation}
The integral corresponding to the defect interaction is much more involved. To make further progress, let us assume $\vec{z}_1=\vec{z}_2=\vec{z}$ (and, with no loss of generality, set $\vec{z}=0$). Note that this means that we are computing correlation functions of defect fields. The integral reduces to

\begin{equation}
\lambda\int |\vec{\Psi}_1|^2\phi^2=\frac{\lambda\nu^2}{256\pi^6G(z_1-z_2)}\int dx^0\int d^{d-1}\vec{x}\frac{1}{(x^0-(z_1^0-z_2^0))^2+\vec{x}^2}\frac{1}{(x^0)^2+\vec{x}^2}
\end{equation}
This can now be done by brute force. Introducing a cut-off so that $|\vec{x}|\in(\epsilon_x,\infty)$, one finds

\begin{equation}
\lambda\int |\vec{\Psi}_1|^2\phi^2=\frac{\lambda\nu^2}{8\pi^2}\log\frac{|z_1^0-z_2^0|}{2\epsilon_x}\,.
\end{equation}
In the following we will assume the appropriate choice of cut-off and simply keep

\begin{equation}
\lambda\int |\vec{\Psi}_1|^2\phi^2=\frac{\lambda\nu^2}{8\pi^2}\log |z_1^0-z_2^0|\,.
\end{equation}

\subsection{Eq.\eqref{d=4defectcorrelator}}
\label{A4}

In this appendix we drive \eqref{d=4defectcorrelator} starting with \eqref{S_eff6d}. For the $\mathcal{O}(\lambda^0)$ terms

\begin{equation}
\frac{\nu}{2}\int \rho_1\delta_T(\vec{x}-\vec{z}_2)+\frac{\nu}{2}\int \rho_2\delta_T(\vec{x}-\vec{z}_1)=-\nu^2\int dx^0 \int\frac{d^{d-1}\vec{p}}{(2\pi)^{d-1}} \frac{e^{-i\vec{p}\cdot(\vec{z}_1-\vec{z}_2)}}{\vec{p}^2}
\end{equation}
Fourier transforming in $d=4-\epsilon$ dimensions we find

\begin{equation}
\frac{\nu}{2}\int \rho_1\delta(\vec{x}-\vec{z}_2)+\frac{\nu}{2}\int \rho_2\delta(\vec{x}-\vec{z}_1)=-\frac{\nu^2}{4\pi}\,\frac{T}{|\vec{z}_1-\vec{z}_2|}(1+\epsilon \log|\vec{z}_1-\vec{z}_2|+\epsilon\frac{\gamma_E+\log(4\pi)}{2}\cdots)\,,
\end{equation}
where $T=\int dx^0$.

In turn, for the $\mathcal{O}(\lambda)$ terms, the first integral is

\begin{equation}
\frac{\lambda}{4}\int \rho_1^3\rho_2=\frac{\lambda\nu^4}{4}\int dx^0 \int\frac{d^{d-1}\vec{p}_1}{(2\pi)^{d-1}}\frac{e^{i\vec{p}_1\cdot(\vec{z}_1-\vec{z}_2)}}{\vec{p}_1^2\,} \int\frac{d^{d-1}\vec{p}_2}{(2\pi)^{d-1}} \int\frac{d^{d-1}\vec{p}_3}{(2\pi)^{d-1}} \frac{1 }{\vec{p}_2^2\,\vec{p}_3^2\,(\vec{p}_1-\vec{p}_2-\vec{p}_3)^2}\,.
\end{equation}
Using eq. 2.8 in \cite{Grozin:2003ak} with $n_1=0$, $n_2=1$, $n_3=1$, $n_4=0$ and $n_5=1$

\begin{equation}
\int\frac{d^{d-1}\vec{p}_2}{(2\pi)^{d-1}} \int\frac{d^{d-1}\vec{p}_3}{(2\pi)^{d-1}} \frac{1 }{\vec{p}_2^2\,\vec{p}_3^2\,(\vec{p}_1-\vec{p}_2-\vec{p}_3)^2}=\frac{\pi^{d-1}}{(2\pi)^{2d-2}}\,(\vec{p}_1^2)^{d-4}\,G(0,1,1,0,1)\,.
\end{equation}
Hence

\begin{equation}
\frac{\lambda}{4}\int \rho_1^3\rho_2=\frac{\lambda\nu^4}{4}\,\frac{\pi^{d-1}}{(2\pi)^{2d-2}}\,G(0,1,1,0,1)\, \int dx^0 \int\frac{d^{d-1}\vec{p}_1}{(2\pi)^{d-1}}\frac{e^{i\vec{p}_1\cdot(\vec{z}_1-\vec{z}_2)}}{(\vec{p}_1^2)^{5-d}}\,.
\end{equation}
Fourier transforming

\begin{equation}
\frac{\lambda}{4}\int \rho_1^3\rho_2=\frac{\lambda\nu^4}{512\pi^3\epsilon}\frac{T}{|\vec{z}_1-\vec{z}_2|}+\frac{3\lambda\nu^4}{512\pi^3}\,\log|\vec{z}_1-\vec{z}_2|\frac{T}{|\vec{z}_1-\vec{z}_2|}+\frac{3\lambda\nu^4}{1024\pi^3}(2+\gamma_E+\log(4\pi))\frac{T}{|\vec{z}_1-\vec{z}_2|}\,.
\end{equation}

As for the remaining integral, it reads

\begin{equation}
\frac{\lambda}{4}\int \rho_1^2\rho_2^2=\frac{\lambda\nu^4}{4}\int dx^0 \int\frac{d^{d-1}\vec{p}_1}{(2\pi)^{d-1}} e^{i\vec{p}_1\cdot(\vec{z_1}-\vec{z}_2)} \int\frac{d^{d-1}\vec{p}_2}{(2\pi)^{d-1}} \int\frac{d^{d-1}\vec{p}_3}{(2\pi)^{d-1}}  \frac{ 1}{(\vec{p}_2+\vec{p}_1)^2\,(\vec{p}_3+\vec{p}_1)^2\,\vec{p}_2^2\,\vec{p}_3^2}\,.
\end{equation}
Using eq. 2.8 in \cite{Grozin:2003ak} with $n_1=1$, $n_2=1$, $n_3=1$, $n_4=1$ and $n_5=0$, and given that $G(1,1,1,1,0)=\pi^3$, this is

\begin{equation}
\frac{\lambda}{4}\int \rho_1^2\rho_2^2=-\frac{\lambda\nu^4}{256}\int dx^0 \int\frac{d^{d-1}\vec{p}_1}{(2\pi)^{d-1}} \frac{e^{i\vec{p}_1\cdot(\vec{z_1}-\vec{z}_2)}}{\vec{p}_1^2}\,.
\end{equation}
Then, Fourier transforming

\begin{equation}
\frac{\lambda}{4}\int \rho_1^2\rho_2^2=-\frac{\lambda\nu^4}{1024\pi }\frac{T}{|\vec{z}_1-\vec{z}_2|}\,.
\end{equation}

\subsection{Eq.\eqref{d=6S}}
\label{A5}

In this appendix we describe the computation of \eqref{d=6S} starting with \eqref{d=6SIntegral}. The $\mathcal{O}(h_2^0)$ part is

\begin{equation}
n\int \frac{\nu}{2} \rho\,\delta_T(x)=-n \frac{\nu^2}{2}\int d\vec{x}_{||}\int\frac{d\vec{p}_T}{(2\pi)^{d_T}} \frac{1}{\vec{p}_T^2}\,.
\end{equation}

In turn, the $\mathcal{O}(h_2)$ term is

\begin{equation}
n\int \frac{h_2}{6}\rho^3=-n\frac{h_2}{6}\nu^3\int d\vec{x}_{||}  \int\frac{d\vec{p}_T}{(2\pi)^{d_T}} \frac{1}{\vec{p}_T^2}\,\int\frac{d\vec{q}_T}{(2\pi)^{d_T}} \frac{ 1}{\vec{q}_T^2\,(\vec{q}_T-\vec{p}_T)^2}   \,.
\end{equation}
The  $\vec{q}_T$ integral is easily done. Introducing Feynman parameters and shifting the integration variable

\begin{equation}
\int\frac{d\vec{q}_T}{(2\pi)^{d_T}} \frac{ 1}{\vec{q}_T^2\,(\vec{q}_T-\vec{p}_T)^2} =\int_0^1 dx \int\frac{d\vec{q}_T}{(2\pi)^{d_T}} \frac{1}{(\vec{q}_T^2+\Delta)^2}\,,\qquad \Delta=x(1-x)\vec{p}_T^2  \,.
\end{equation}
The integral is easily doable. Writing now $d_T=d-2$, in $d=6-\epsilon$ dimensions we have

\begin{equation}
\int\frac{d\vec{q}_T}{(2\pi)^{d_T}} \frac{ 1}{\vec{q}_T^2\,(\vec{q}_T-\vec{p}_T)^2} =\frac{2\pi^{\frac{d_{T}}{2}}}{\Gamma(\frac{d_{T}}{2})} \frac{1}{(2\pi)^{d_T}} \frac{2-d_T}{4}\frac{\pi}{\sin(\frac{d_T\pi}{2})}  \int_0^1 dx\, \Delta^{\frac{d_T}{2}-2}= \frac{1}{8\pi^2\,\epsilon} -\frac{\log|\vec{p}_T|}{8\pi^2}\,.
\end{equation}
(here $\log|\vec{p}_T|^2=\log(e^{-2+\gamma_E-\log(4\pi)}\mu|\vec{p}_T|^2)$).

\subsection{Eq.\eqref{d=6correlator}}
\label{A6}

In this appendix we describe the evaluation of the integrals in the exponent in \eqref{d=6Correlator} leading to \eqref{d=6correlator}. The contribution of $S_{\rm eff}^C$ can be borrowed from \cite{Arias-Tamargo:2020fow}. As for the first contribution in the remaining terms

\begin{equation}
\int \rho_D |\Phi_C|^2=-\frac{\nu}{(4\pi^2)(4\pi^3)^2\,G(z_1-z_2)}\int dx \frac{1}{|x-z_1|^4\,|x-z_2|^4\,|\vec{x}_T|^2}\,.
\end{equation}
To make further progress, we will assume that $\vec{z}_{iT}=0$, so that we are computing correlators of defect fields. Then

\begin{equation}
\int \rho_D |\Phi_C|^2=-\frac{\nu}{(4\pi^2)(4\pi^3)^2\,G(z_1-z_2)}\int d\vec{x}_T\int d\vec{x}_{||} \frac{1}{((\vec{x}_{||}-\vec{z}_{||})^2+\vec{x}_{T}^2)^2}\frac{1}{(\vec{x}_{||}^2+\vec{x}_{T}^2)^2} \frac{1}{\vec{x}_T^2}\,,
\end{equation}
where $\vec{z}_{||}=\vec{z}_{1||}-\vec{z}_{2||}$. The integral is (that the same remarks concerning the regularization, using a cut-off in position space, of the integrals apply as in the case in appendix \ref{A3}) 

\begin{equation}
\int d\vec{x}_T\int d\vec{x}_{||} \frac{1}{((\vec{x}_{||}-\vec{z}_{||})^2+\vec{x}_{T}^2)^2}\frac{1}{(\vec{x}_{||}^2+\vec{x}_{T}^2)^2} \frac{1}{\vec{x}_T^2}=\frac{\pi^3}{z^4}\log|\vec{z}_{||}|\,.
\end{equation}
Therefore

\begin{equation}
\frac{h_1n}{2}\int \rho_D |\Phi_C|^2=-n\frac{h_1\nu}{32\pi^2}\log|\vec{z}_{1||}-\vec{z}_{2||}|\,.
\end{equation}

As for the second contribution, it reads 

\begin{equation}
\int \rho_C \delta_T=-\frac{h_1}{G(z_1-z_2)}\int dx \int dy\, G(x-y)G(y-z_1)G(y-z_2) \delta_T(x)\,.
\end{equation}
This can be re-arranged as follows

\begin{equation}
\int \rho_C \delta_T=\frac{h_1}{\nu} \int dy\, \Big(-\nu\int dxG(y-x)\delta_T(x)\Big)\,\frac{G(y-z_1)}{\sqrt{G(z_1-z_2)}}\,\frac{G(y-z_2)}{\sqrt{G(z_1-z_2)}} \,.
\end{equation}
We recognize here $\rho_D$, $\Phi_C$ and $\Phi_C^{\star}$. Thus

\begin{equation}
\int \rho_C \delta_T=\frac{h_1}{\nu} \int \rho_D\,|\Phi_C|^2\,.
\end{equation}
Hence, borrowing the previous results

\begin{equation}
n\nu \int \rho_C \delta_T= - \frac{nh_1\nu}{16\pi^2}\log|\vec{z}_{1||}-\vec{z}_{2||}|\,.
\end{equation}

\subsection{Eq.\eqref{d=6DefectCorrelatorResult}}
\label{A7}

In this appendix we give details of the computation of \eqref{d=6DefectCorrelatorResult} starting with the integral in \eqref{d=6DefectCorrelatorIntegral}. The $\mathcal{O}(h_2^0)$ term in \eqref{d=6DefectCorrelatorIntegral}

\begin{equation}
\nu\int dx \rho_2\,\delta_T(\vec{z}_{1T})=-\nu^2 \int d\vec{x}_{||} \int \frac{d^{d_T}\vec{p}_T}{(2\pi)^{d_T}}\,\frac{e^{i\vec{p}_T\cdot(\vec{z}_{1T}-\vec{z}_{2T})}}{\vec{p}_T^2} 
\end{equation}
Doing the Fourier transform

\begin{equation}
\nu\int dx \rho_2\,\delta_T(\vec{z}_{1T})=-\nu^2 V\Big(\frac{1}{4\pi^2\vec{z}_T^2}+\frac{\epsilon}{8\pi^2\vec{z}_T^2} (\gamma_E+\log\pi)+\frac{\epsilon}{4\pi^2}\frac{\log|\vec{z}_T|}{\vec{z}_T^2}\Big)\,.
\end{equation}

The $\mathcal{O}(h_2)$ term is

\begin{equation}
\int dx\rho_1^2\rho_2=-\nu^3V \int \frac{d^{d_T}\vec{p}_T}{(2\pi)^{d_T}}  \frac{e^{i(\vec{p}_T+\vec{q}_T)\cdot(\vec{z}_{1T}-\vec{z}_{2T})}}{\vec{p}_T^2}\int \frac{d^{d_T}\vec{q}_T}{(2\pi)^{d_T}} \frac{1 }{\vec{q}_T^2(\vec{p}_T+\vec{q}_T)^2}\,.
\end{equation}
Doing the $\vec{q}_T$ integral 

\begin{eqnarray}
\int dx\rho_1^2\rho_2&=&-\frac{\nu^3V}{8\pi^2\epsilon} \int \frac{d^{d_T}\vec{p}_T}{(2\pi)^{d_T}} e^{i(\vec{p}_T+\vec{q}_T)\cdot(\vec{z}_{1T}-\vec{z}_{2T})} (\frac{1}{\vec{p}_T^2}-\epsilon \frac{\log|\vec{p}_T|}{\vec{p}_T^2})\nonumber \\ && -\frac{\nu^3V}{16\pi^2}(2-\gamma_E+\log(4\pi)) \int \frac{d^{d_T}\vec{p}_T}{(2\pi)^{d_T}}  \frac{e^{i(\vec{p}_T+\vec{q}_T)\cdot(\vec{z}_{1T}-\vec{z}_{2T})}}{\vec{p}_T^2} \,.
\end{eqnarray}
This can be regarded as the expansion of

\begin{eqnarray}
\int dx\rho_1^2\rho_2&=&-\frac{\nu^3V}{8\pi^2\epsilon} \int \frac{d^{d_T}\vec{p}_T}{(2\pi)^{d_T}} e^{i(\vec{p}_T+\vec{q}_T)\cdot(\vec{z}_{1T}-\vec{z}_{2T})} \frac{1}{(\vec{p}_T^2)^{1+\frac{\epsilon}{2}}}\nonumber \\ && -\frac{\nu^3V}{16\pi^2}(2-\gamma_E+\log(4\pi)) \int \frac{d^{d_T}\vec{p}_T}{(2\pi)^{d_T}}  \frac{e^{i(\vec{p}_T+\vec{q}_T)\cdot(\vec{z}_{1T}-\vec{z}_{2T})}}{\vec{p}_T^2} \,.
\end{eqnarray}
Fourier-transforming\footnote{I would like to thank the referee for spotting a mistake in the numerical coefficient of the last term in the first line which lead to an incorrect result for the correlator in eq.\eqref{d=6DefectCorrelatorResult} in a previous version of this paper.}

\begin{eqnarray}
\int dx\rho_1^2\rho_2&=&-\frac{\nu^3V}{8\pi^2\epsilon} \Big( \frac{1}{4\pi^2\vec{z}_T^2} +\epsilon \frac{3\gamma_E+\log\frac{\pi}{4}}{8\pi^2\vec{z}_T^2} +\frac{\epsilon}{2\pi^2} \frac{\log|\vec{z}_T|}{\vec{z}_T^2}\Big)\nonumber \\ && -\frac{\nu^3V}{16\pi^2}(2-\gamma_E+\log(4\pi)) \Big(\frac{1}{4\pi^2|\vec{z}_T|^2} + \frac{\epsilon}{4\pi^2} \frac{\log|\vec{z}_T|}{\vec{z}_T^2} + \epsilon\frac{\gamma_E+\log\pi}{8\pi^2\vec{z}_T^2}\Big) \,.
\end{eqnarray}

\end{appendix}

\end{document}